# Identification of higher-order continua equivalent to a Cauchy elastic composite☆

A. Bacigalupo[a], M. Paggi[a], F. Dal Corso[b], D. Bigoni[*,b]

[a]*IMT School for Advanced Studies Lucca, Piazza San Francesco 19, 55100 Lucca, Italy*
[b]*DICAM, University of Trento, Via Mesiano 77, 38123 Trento, Italy*

## Abstract

A heterogeneous Cauchy elastic material may display micromechanical effects that can be modeled in a homogeneous equivalent material through the introduction of higher-order elastic continua. Asymptotic homogenization techniques provide an elegant and rigorous route to the evaluation of equivalent higher-order materials, but are often of difficult and awkward practical implementation. On the other hand, identification techniques, though relying on simplifying assumptions, are of straightforward use. A novel strategy for the identification of equivalent second-gradient Mindlin solids is proposed in an attempt to combine the accuracy of asymptotic techniques with the simplicity of identification approaches. Following the asymptotic homogenization scheme, the overall behaviour is defined via perturbation functions, which (differently from the asymptotic scheme) are evaluated on a finite domain obtained as the periodic repetition of cells and subject to quadratic displacement boundary conditions. As a consequence, the periodicity of the perturbation function is satisfied only in an approximate sense, nevertheless results from the proposed identification algorithm are shown to be reasonably accurate.

## 1. Introduction

At a first-level of approximation, homogenization techniques allow to represent a Cauchy elastic material, heterogeneous at the microscale, as a homogeneous equivalent Cauchy elastic material at the macroscale. Although of primary importance, the first-level approximation fails to describe fine mechanical responses in which the heterogeneity at the microscale introduces size-dependency. This dependency can be enforced by assuming that at the macroscale the material behaves as a higher-order elastic continuum, a constitutive model thoroughly considered by G.A. Maugin [37, 39, 40, 41] to whom this article is dedicated (the reader is also referred to [32, 33, 44] for other relevant contributions on micromorphic and higher-order elastic continua).

In this way, the length scales of the higher-order continuum are not postulated, but determined through the identification procedure, so that they become related to the microstructural geometry of the heterogeneous Cauchy material. The schemes proposed so far for attacking this problem can be distinguished in three categories, namely, asymptotic homogenization approaches [2, 3, 11, 15, 16, 17, 19, 20, 21, 25, 29, 42, 46, 50], variational asymptotic schemes [9, 12, 13, 14, 47, 48, 49, 53] and several identification techniques, including analytical [5, 6, 7, 10, 18, 38] and computational approaches [1, 8, 24, 26, 27, 28, 30, 34, 35, 36, 45, 51, 52, 54]. The asymptotic approaches are elegant, rigorous and have been shown to yield very accurate results, but they are complex and often of very difficult implementation. On the other hand, identification techniques are based on simplifying assumptions and therefore are of easy implementation, but their accuracy is reduced when compared to that of the asymptotic approaches.

In the present article a new identification procedure is proposed in which the perturbation functions, introduced in the asymptotic homogenization scheme, are numerically evaluated on a finite domain, which is realized as the periodic repetition of a rectangular cell and is subject to imposed quadratic displacement at its external boundary. In this procedure, the periodicity of the perturbation function is satisfied only in an approximate sense so that the proposed approach results easier to be implemented than the asymptotic scheme. Although the simplification introduces an approximation, results from the proposed scheme are shown to nicely predict the behavior of the heterogeneous material and to perform better than other identification procedures (for instance the second-order computational two-scales homogenization, which often predicts higher-order effects even in the homogeneous material limit case).

---

## 2. Theoretical identification strategy

*2.1. Formulation of the kinematics at the micro and macro scales*

A heterogeneous Cauchy material having a periodic microstructure is considered at the *microscale*. By restricting the attention for the sake of simplicity to a planar problem, the periodic rectangular cell $\mathcal{A}$ (Fig.1, lower part) is repeated within the plane to realize the *finite* domain $\mathcal{L}$ occupied by the elastic solid (Fig.1, upper part),

$$\mathcal{A} = \left[-\frac{\epsilon}{2}, \frac{\epsilon}{2}\right] \times \left[-r\frac{\epsilon}{2}, r\frac{\epsilon}{2}\right],$$
$$\mathcal{L} = \left[-n_1\frac{\epsilon}{2}, n_1\frac{\epsilon}{2}\right] \times \left[-rn_2\frac{\epsilon}{2}, rn_2\frac{\epsilon}{2}\right]. \quad (1)$$

The parameter $\epsilon$ denotes the characteristic size of the periodic cell, while $r$ rules its aspect ratio ($r \in \mathbb{R}^+$), and $n_1$ and $n_2 \in \mathbb{N}$ define the repetition of the cell along the Cartesian axes $x_1$ and $x_2$, respectively. Being $\mathbf{v}_1$ and $\mathbf{v}_2$ the periodicity vectors,

$$\mathbf{v}_1 = \epsilon \begin{bmatrix} 1 \\ 0 \end{bmatrix}, \qquad \mathbf{v}_2 = r\epsilon \begin{bmatrix} 0 \\ 1 \end{bmatrix}, \quad (2)$$

the *microstructural* elasticity tensor $\mathbf{C}^{m,\epsilon}(\mathbf{x})$, with components $C^{m,\epsilon}_{ijhk}(\mathbf{x})$, satisfies the following periodicity property

$$C^{m,\epsilon}_{pqst}(\mathbf{x}) = C^{m,\epsilon}_{pqst}(\mathbf{x} + i\mathbf{v}_1 + j\mathbf{v}_2), \qquad i,j \in \mathbb{Z}, \quad (3)$$
$$\forall \mathbf{x} \text{ and } \mathbf{x} + i\mathbf{v}_1 + j\mathbf{v}_2 \in \mathcal{L}.$$

Equilibrium for the heterogeneous Cauchy material in the presence of body forces $\mathbf{f}(\mathbf{x})$ implies the following partial differential equation for the *micro-displacement* field $\mathbf{u}(\mathbf{x})$:

$$\frac{\partial}{\partial x_j}\left(C^{m,\epsilon}_{ijhk}(\mathbf{x})\frac{\partial u_h(\mathbf{x})}{\partial x_k}\right) + f_i(\mathbf{x}) = 0, \qquad \mathbf{x} \in \mathcal{L}. \quad (4)$$

The equilibrium equations are complemented by boundary conditions applied over the boundary of $\mathcal{L}$ which are considered as the following generic quadratic functions of the coordinates $\mathbf{x} \in \partial\mathcal{L}$ [5, 8, 27]:

$$u_h(\mathbf{x}) = \zeta_h + \alpha_{hp}x_p + \frac{1}{2}\beta_{hpq}x_p x_q, \quad h,p,q = 1,2, \quad (5)$$
$$\mathbf{x} \in \partial\mathcal{L},$$

where $\zeta_h$, $\alpha_{hp}$, and $\beta_{hpq}$ are the components of the first, second, and third-order tensors $\boldsymbol{\zeta}$, $\boldsymbol{\alpha}$, and $\boldsymbol{\beta}$, respectively. Note that $\boldsymbol{\zeta}$, as well as the skew-symmetric part of $\boldsymbol{\alpha}$, do not affect the mechanical response of the composite since they correspond to a rigid-body motion.

At the *macroscale*, an equivalent homogenous solid is modeled as a second-gradient homogenous material [32, 44], which is characterized by the unknown fourth-, fifth-, and sixth- order constitutive tensors $C_{ijhk}$, $Y_{ijhks}$, and $S_{ijrhks}$, defined on the same domain $\mathcal{L}$ where the periodic

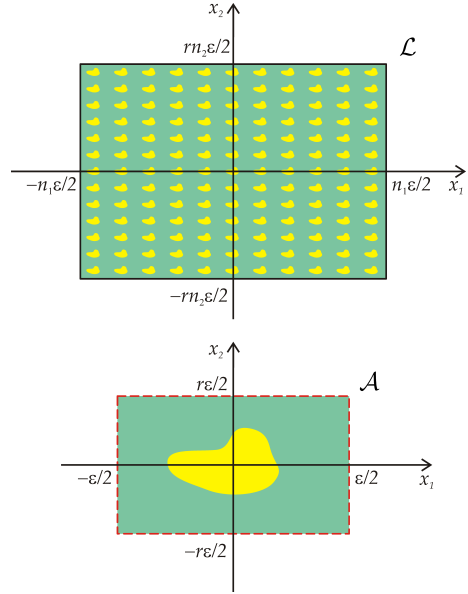

**Figure 1:** The periodic composite material (upper part) occupying the *finite* domain $\mathcal{L}$ is obtained as the periodic repetition of the rectangular cell $\mathcal{A}$ (lower part) along the axes $x_1$ and $x_2$ for $n_1$ and $n_2$ times, respectively. The rectangular cell $\mathcal{A}$ is defined by the characteristic dimension $\epsilon$ and the aspect ratio parameter $r$. The phases composing the rectangular cell are modelled as Cauchy elastic materials.

Cauchy material is defined. In the presence of the body force $\mathbf{f}(\mathbf{x})$, the equilibrium for the equivalent material is governed by the following partial differential equation for the *macro-displacement* field $\mathbf{U}(\mathbf{x})$:

$$S_{ijrhks}\frac{\partial^4 U_h(\mathbf{x})}{\partial x_j \partial x_r \partial x_k \partial x_s} + Y_{ijhks}\frac{\partial^3 U_h(\mathbf{x})}{\partial x_j \partial x_k \partial x_s}$$
$$-C_{ijhk}\frac{\partial^2 U_h(\mathbf{x})}{\partial x_j \partial x_k} - f_i(\mathbf{x}) = 0, \qquad \mathbf{x} \in \mathcal{L}. \quad (6)$$

Dirichlet boundary conditions (the same applied to the heterogeneous material) are imposed, see Eq.(5), complemented by Neumann boundary conditions, related to the derivatives normal to the boundary [5]

$$\begin{cases} U_h(\mathbf{x}) = u_h(\mathbf{x}), & \mathbf{x} \in \partial\mathcal{L}, \\ \dfrac{\partial U_h(\mathbf{x})}{\partial x_j}n_j = (\alpha_{hj} + \beta_{hpj}x_p)n_j, & \mathbf{x} \in \partial\mathcal{L}. \end{cases} \quad (7)$$

The identification of the constitutive tensors of the second-order homogenized material is a complex task. To simplify this procedure, a homogenization approach is proposed by assuming the following ansatz: *the macro-displacement field is a quadratic function of the position vector*, namely, Eq.(7)$_1$ is analytically extended within the domain $\mathcal{L}$:

$$U_h(\mathbf{x}) = \zeta_h + \alpha_{hp}x_p + \frac{1}{2}\beta_{hpq}x_p x_q, \qquad \mathbf{x} \in \mathcal{L}. \quad (8)$$

The accuracy of this assumption (which restricts the field of the admissible solutions) will be tested against the direct solution of the microscale problem in the benchmark tests reported in Section 4. By inserting the quadratic ansatz (8) into the equilibrium equations (6) under the hypothesis of vanishing body-forces, $f_i = 0$, the following condition on the admissible subset of curvature parameters $\beta_{hkj}$ is derived:

$$C_{ijhk}\beta_{hkj} = 0, \tag{9}$$

which, without any loss of generality, for a centro-symmetric microstructure it is equivalently expressed by

$$\beta_{hkh} = \frac{C_{kkkk}}{C_{hhkk}+C_{hkhk}}\beta_{kkk} - \frac{C_{hkhk}}{C_{hhkk}+C_{hkhk}}\beta_{khh},$$
$$h, k \text{ not summed}. \tag{10}$$

Based on the previous assumption, the kinematic field solution of the macro-scale problem will assume the form (8), subject to the constraint (9). The corresponding expressions of the macro-strain and second gradient of the macro-displacement are:

$$H_{ip}(\mathbf{x}) = \frac{\partial U_i(\mathbf{x})}{\partial x_p} = \alpha_{ip} + \beta_{ipq}x_q,$$
$$\kappa_{ipq}(\mathbf{x}) = \frac{\partial^2 U_i(\mathbf{x})}{\partial x_p \partial x_q} = \beta_{ipq}. \tag{11}$$

### 2.2. Down-scaling and periodic perturbation functions

Introducing the microscopic (or fast) variable $\boldsymbol{\xi} = \mathbf{x}/\epsilon$ [11, 17, 48], which defines the coordinate within the unit-cell $\mathcal{Q} = [-1/2, 1/2] \times [-r/2, r/2]$, the constitutive elastic tensor of the heterogeneous solid is mapped as

$$C_{ijhk}^{m,\epsilon}(\mathbf{x}) = C_{ijhk}^{m}\left(\boldsymbol{\xi} = \frac{\mathbf{x}}{\epsilon}\right). \tag{12}$$

A substitution of the mapping (12) in the equilibrium equation (4) defines the differential problem governing the *micro-displacement* field $\mathbf{u}(\mathbf{x}, \boldsymbol{\xi} = \mathbf{x}/\epsilon)$.

Similarly to the standard asymptotic technique used for periodic infinite materials subject to body-forces (with periodicity larger than the cell size), the generic *micro-field* $\mathbf{v}(\mathbf{x}, \boldsymbol{\xi} = \mathbf{x}/\epsilon)$ is asymptotically expressed in a power series of the characteristic size parameter $\epsilon$ [11, 17, 48]. The terms of this series depend on the *macro-spatial* derivatives of the *macro-field* $\mathbf{V}(\mathbf{x})$ and on the perturbation functions $\mathbf{N}^{(n)}$ ($n = 1, 2, 3, \ldots$). A truncation of the asymptotic representation at the third power in $\epsilon$, yields the down-scaling relation

$$v_h\left(\mathbf{x}, \frac{\mathbf{x}}{\epsilon}\right) \approx V_h(\mathbf{x}) + \epsilon N_{hst}^{(1)}\left(\frac{\mathbf{x}}{\epsilon}\right)\frac{\partial V_s(\mathbf{x})}{\partial x_t}$$
$$+ \epsilon^2 N_{hqst}^{(2)}\left(\frac{\mathbf{x}}{\epsilon}\right)\frac{\partial^2 V_q(\mathbf{x})}{\partial x_s \partial x_t}, \tag{13}$$

where $\mathbf{N}^{(1)}$ and $\mathbf{N}^{(2)}$ are the first- and second- order perturbation functions of the fast variable $\boldsymbol{\xi} = \mathbf{x}/\epsilon$. Note that in the case of large values of both $n_1$ and $n_2$ the perturbation functions tend to be $\mathcal{A}$-periodic.

A substitution of the representation $\mathbf{V}(\mathbf{x}) = \mathbf{U}(\mathbf{x})$, $\mathbf{v}(\mathbf{x}, \mathbf{x}/\epsilon) = \mathbf{u}(\mathbf{x}, \mathbf{x}/\epsilon)$ in Eq.(13), with $\mathbf{U}$, Eqs.(8), satisfying the constraint (9), yields the down-scaling relation for the micro-displacement field $\mathbf{u}(\mathbf{x}, \mathbf{x}/\epsilon)$ as

$$u_h(\mathbf{x}, \boldsymbol{\xi} = \mathbf{x}/\epsilon) \approx \left\{\zeta_h + \left[\delta_{hs}x_t + \epsilon N_{hst}^{(1)}(\boldsymbol{\xi})\right]\alpha_{st}\right.$$
$$+ \left[\frac{1}{2}\delta_{hq}x_s x_t + \epsilon N_{hqs}^{(1)}(\boldsymbol{\xi})x_t\right.$$
$$\left.\left. + \epsilon^2 N_{hqst}^{(2)}(\boldsymbol{\xi})\right]\beta_{qst}\right\}\bigg|_{\boldsymbol{\xi}=\mathbf{x}/\epsilon}, \tag{14}$$

where the parameters $\beta_{qst}$, restricted to satisfy the kinematical constraint (10), are given by the linear relation

$$\beta_{qst} = M_{qstij}\widehat{\beta}_{ij}. \tag{15}$$

In Eq.(15), $\widehat{\beta}_{ij}$ are the components of the second-order tensor $\widehat{\boldsymbol{\beta}}$, collecting the independent 'curvature parameters', defined as

$$\widehat{\beta}_{ij} = \beta_{ijj}, \qquad j \text{ not summed}, \tag{16}$$

while $M_{qstij}$ are the components of the *transformation* fifth-order tensor $\mathbf{M}$, possessing the following non-null components

$$M_{11111} = M_{22222} = M_{12212} = M_{21121} = 1,$$
$$M_{12122} = M_{11222} = -\frac{C_{2222}}{C_{1122}+C_{1212}},$$
$$M_{12121} = M_{11221} = M_{21212} = M_{22112} = -\frac{C_{1212}}{C_{1122}+C_{1212}},$$
$$M_{21211} = M_{22111} = -\frac{C_{1111}}{C_{1122}+C_{1212}}. \tag{17}$$

Keeping into account Eq. (15), the down-scaling relation (14) can be rewritten in terms of the independent parameters $\widehat{\beta}_{ij}$ as

$$u_h(\mathbf{x}, \boldsymbol{\xi} = \mathbf{x}/\epsilon) \approx \left\{\zeta_h + \left[\delta_{hs}x_t + \epsilon N_{hst}^{(1)}(\boldsymbol{\xi})\right]\alpha_{st}\right.$$
$$+ \left[\frac{1}{2}M_{hstij}x_s x_t + \epsilon \widehat{N}_{htij}^{(1)}(\boldsymbol{\xi})x_t\right.$$
$$\left.\left. + \epsilon^2 \widehat{N}_{hij}^{(2)}(\boldsymbol{\xi})\right]\widehat{\beta}_{ij}\right\}\bigg|_{\boldsymbol{\xi}=\mathbf{x}/\epsilon}, \tag{18}$$

where the condensed perturbation functions $\widehat{\mathbf{N}}^{(1)}$ and $\widehat{\mathbf{N}}^{(2)}$ are introduced, with components

$$\widehat{N}_{htij}^{(1)}(\boldsymbol{\xi}) = N_{hqs}^{(1)}(\boldsymbol{\xi})M_{qstij}, \tag{19a}$$
$$\widehat{N}_{hij}^{(2)}(\boldsymbol{\xi}) = N_{hqst}^{(2)}(\boldsymbol{\xi})M_{qstij}. \tag{19b}$$

## 2.3. Equivalent constitutive parameters through energy match

Through the micro/macro scale-separation, the averaged micro-scale strain energy $\mathcal{E}_m$ can be expressed as [11, 48, 53]

$$\mathcal{E}_m \doteq \int_{\mathcal{L}} \left[ \int_Q \phi_m(\mathbf{x}, \boldsymbol{\xi}) \, d\boldsymbol{\xi} \right] d\mathbf{x}, \qquad (20)$$

where the strain energy density $\phi_m$ is given by

$$\phi_m(\mathbf{x}, \boldsymbol{\xi}) = \frac{1}{2} C^m_{ijhk}(\boldsymbol{\xi})[\nabla u_i(\mathbf{x}, \boldsymbol{\xi})]_j [\nabla u_h(\mathbf{x}, \boldsymbol{\xi})]_k \qquad (21)$$

and the differential operator $[\nabla u_i(\mathbf{x}, \boldsymbol{\xi})]_j$ is defined in terms of differentiation with respect of both the slow variable $\mathbf{x}$ and the fast variable $\boldsymbol{\xi}$ as follows

$$[\nabla u_i(\mathbf{x}, \boldsymbol{\xi})]_j = \frac{\partial u_i}{\partial x_j} + \frac{1}{\epsilon} \frac{\partial u_i}{\partial \xi_j}. \qquad (22)$$

The differential operator applied to the micro-displacement field $\mathbf{u}$, defined by Eq.(13), provides the following relation

$$[\nabla u_i(\mathbf{x}, \boldsymbol{\xi})]_j = \left( \delta_{is}\delta_{jt} + \frac{\partial N^{(1)}_{ist}(\boldsymbol{\xi})}{\partial \xi_j} \right) H_{st}(\mathbf{x})$$
$$+ \epsilon \left( N^{(1)}_{iqs}(\boldsymbol{\xi})\delta_{tj} + \frac{\partial N^{(2)}_{iqst}(\boldsymbol{\xi})}{\partial \xi_j} \right) \kappa_{qst}(\mathbf{x}), \qquad (23)$$

involving the macro-strain $\mathbf{H}(\mathbf{x})$ and the second-gradient of macro-displacement $\boldsymbol{\kappa}(\mathbf{x})$, Eq.(11), the latter restricted to self-equilibrated fields, Eq.(10), so that

$$\kappa_{qst} = M_{qstij}\widehat{\kappa}_{ij}, \qquad (24)$$

where $\widehat{\kappa}_{ij} = \kappa_{ijj}$ ($j$ not summed).

Considering the relation (23), the energy density at the micro-level (20) can be expressed as

$$\phi_m(\mathbf{x}, \boldsymbol{\xi}) = \frac{1}{2} \{ \mathcal{C}_{ijhk}(\boldsymbol{\xi}) H_{ij}(\mathbf{x}) H_{hk}(\mathbf{x})$$
$$+ \epsilon \mathcal{Y}_{ijhkl}(\boldsymbol{\xi})[H_{ij}(\mathbf{x})\kappa_{hkl}(\mathbf{x}) + \kappa_{ijh}(\mathbf{x})H_{kl}(\mathbf{x})]$$
$$+ \epsilon^2 \mathcal{S}_{ijhklm}(\boldsymbol{\xi})\kappa_{ijh}(\mathbf{x})\kappa_{klm}(\mathbf{x}) \}, \qquad (25)$$

where the components of the fourth-, fifth-, and sixth- order tensors $\mathcal{C}, \mathcal{Y}, \mathcal{S}$, can be expressed in terms of the elasticity tensor at the micro-scale $\mathbf{C}^m$ and the perturbation functions $\mathbf{N}^{(1)}$ and $\mathbf{N}^{(2)}$ as

$$\mathcal{C}_{ijhk}(\boldsymbol{\xi}) = C^m_{pqst}(\boldsymbol{\xi}) \left( \delta_{ip}\delta_{jq} + \frac{\partial N^{(1)}_{pij}(\boldsymbol{\xi})}{\partial \xi_q} \right)$$
$$\times \left( \delta_{hs}\delta_{kt} + \frac{\partial N^{(1)}_{shk}(\boldsymbol{\xi})}{\partial \xi_k} \right),$$

$$\mathcal{Y}_{ijhkl}(\boldsymbol{\xi}) = C^m_{pqst}(\boldsymbol{\xi}) \left( \delta_{ip}\delta_{jq} + \frac{\partial N^{(1)}_{pij}(\boldsymbol{\xi})}{\partial \xi_q} \right)$$
$$\times \left( \delta_{lt} N^{(1)}_{shk}(\boldsymbol{\xi}) + \frac{\partial N^{(2)}_{shkl}(\boldsymbol{\xi})}{\partial \xi_t} \right),$$

$$\mathcal{S}_{ijhklm}(\boldsymbol{\xi}) = C^m_{pqst}(\boldsymbol{\xi}) \left( \delta_{hq} N^{(1)}_{pij}(\boldsymbol{\xi}) + \frac{\partial N^{(2)}_{pijh}(\boldsymbol{\xi})}{\partial \xi_q} \right)$$
$$\times \left( \delta_{mt} N^{(1)}_{skl}(\boldsymbol{\xi}) + \frac{\partial N^{(2)}_{sklm}(\boldsymbol{\xi})}{\partial \xi_t} \right). \qquad (26)$$

Recalling that the considered boundary conditions satisfy the kinematical constraint (15), the energy density $\phi_m$ (25) can be reduced to the following quadratic function of the second-order tensors $\mathbf{H}$ and $\widehat{\boldsymbol{\kappa}}$

$$\phi_m(\mathbf{x}, \boldsymbol{\xi}) = \frac{1}{2} \{ \mathcal{C}_{ijhk}(\boldsymbol{\xi}) H_{ij}(\mathbf{x}) H_{hk}(\mathbf{x})$$
$$+ \epsilon \widehat{\mathcal{Y}}_{ijhk}(\boldsymbol{\xi})[H_{ij}(\mathbf{x})\widehat{\kappa}_{hk}(\mathbf{x}) + \widehat{\kappa}_{ij}(\mathbf{x})H_{hk}(\mathbf{x})]$$
$$+ \epsilon^2 \widehat{\mathcal{S}}_{ijhk}(\boldsymbol{\xi})\widehat{\kappa}_{ij}(\mathbf{x})\widehat{\kappa}_{hk}(\mathbf{x}) \} \qquad (27)$$

where the fourth-order tensors $\widehat{\mathcal{Y}}$ and $\widehat{\mathcal{S}}$ have been introduced, which represents *the condensed representation* for the stiffness tensors of the equivalent higher-order material. These tensors are related to the fifth- and sixth- order tensors $\mathcal{Y}$ and $\mathcal{S}$ through the following relations

$$\widehat{\mathcal{Y}}_{ijhk}(\boldsymbol{\xi}) = \mathcal{Y}_{ijrst}(\boldsymbol{\xi}) M_{rsthk},$$
$$\widehat{\mathcal{S}}_{ijhk}(\boldsymbol{\xi}) = \mathcal{S}_{lmnrst}(\boldsymbol{\xi}) M_{lmnij} M_{rsthk}, \qquad (28)$$

so that Eqs.(19a) and (19b) yield

$$\widehat{\mathcal{Y}}_{ijhk}(\boldsymbol{\xi}) = C^m_{pqst}(\boldsymbol{\xi}) \left( \delta_{ip}\delta_{jq} + \frac{\partial N^{(1)}_{pij}(\boldsymbol{\xi})}{\partial \xi_q} \right)$$
$$\times \left( \widehat{N}^{(1)}_{sthk}(\boldsymbol{\xi}) + \frac{\partial \widehat{N}^{(2)}_{shk}(\boldsymbol{\xi})}{\partial \xi_t} \right),$$

$$\widehat{\mathcal{S}}_{ijhk}(\boldsymbol{\xi}) = C^m_{pqst}(\boldsymbol{\xi}) \left( \widehat{N}^{(1)}_{pqij}(\boldsymbol{\xi}) + \frac{\partial \widehat{N}^{(2)}_{pij}(\boldsymbol{\xi})}{\partial \xi_q} \right)$$
$$\times \left( \widehat{N}^{(1)}_{sthk}(\boldsymbol{\xi}) + \frac{\partial \widehat{N}^{(2)}_{shk}(\boldsymbol{\xi})}{\partial \xi_t} \right). \qquad (29)$$

With reference to the subset of curvature parameters $\widehat{\kappa}_{ij}$ imposed through the kinematical constraint (10), the energy $\mathcal{E}_M$ stored in the equivalent second-gradient homo-

geneous material is

$$\mathcal{E}_M = \int_\mathcal{L} \phi_M(\mathbf{x}) d\mathbf{x} \qquad (30)$$

where the strain energy density at the macro-level is defined as

$$\phi_M(\mathbf{x}) = \frac{1}{2}\left[ C_{ijhk} H_{ij} H_{hk} + \widehat{Y}_{ijhk}(H_{ij}\widehat{\kappa}_{hk} + \widehat{\kappa}_{ij} H_{hk}) \right. $$
$$\left. + \widehat{S}_{ijhk}\widehat{\kappa}_{ij}\widehat{\kappa}_{hk} \right], \qquad (31)$$

being $\mathbf{C}$, $\widehat{\mathbf{Y}}$, and $\widehat{\mathbf{S}}$ the fourth-order constitutive tensors. Note that in the stored energy, the condensed stiffnesses $\widehat{\mathbf{Y}}$ and $\widehat{\mathbf{S}}$ are present as a consequence of the kinematical assumption (10), and have reduced orders with respect to the stiffness tensors $\mathbf{Y}$ and $\mathbf{S}$.

A comparison between the energy stored in the equivalent material, Eq.(30), and that stored in the heterogeneous material, Eq.(25), leads to the following equivalent stiffness for the (*condensed*) equivalent second-gradient homogeneous material

$$C_{ijhk} = \int_Q \mathcal{C}_{ijhk}(\boldsymbol{\xi}) d\boldsymbol{\xi},$$
$$\widehat{Y}_{ijhk} = \epsilon \int_Q \widehat{\mathcal{Y}}_{ijhk}(\boldsymbol{\xi}) d\boldsymbol{\xi}, \qquad (32)$$
$$\widehat{S}_{ijhk} = \epsilon^2 \int_Q \widehat{\mathcal{S}}_{ijhk}(\boldsymbol{\xi}) d\boldsymbol{\xi}.$$

The obtained equivalent *condensed* constitutive tensors (32) depend only on the geometrical and mechanical properties through the elasticity tensor $\mathbf{C}^m(\boldsymbol{\xi})$, which defines the periodic heterogeneous material, and on the perturbation functions $\mathbf{N}^{(1)}(\boldsymbol{\xi})$ and $\widehat{\mathbf{N}}^{(2)}(\boldsymbol{\xi})$ (obtained from the unit cell problem) and $\widehat{\mathbf{N}}^{(1)}(\boldsymbol{\xi})$, Eq.(19a). It is worth remarking that the present homogenization scheme implies that the following properties (not always satisfied in other homogenization procedures) for the tensors $\widehat{\mathbf{Y}}$ and $\widehat{\mathbf{S}}$ hold true:

- a centro-symmetric microstructure provides a null condensed tensor $\widehat{\mathbf{Y}} = \mathbf{0}$;

- in the limit when the microstructure disappears and the material becomes homogeneous, the condensed tensors vanish, $\widehat{\mathbf{Y}} = \widehat{\mathbf{S}} = \mathbf{0}$, due to the annihilation of the perturbation functions $\mathbf{N}^{(1)} = \widehat{\mathbf{N}}^{(2)} = \mathbf{0}$;

- a positive definite condensed tensor $\widehat{\mathbf{S}}$ is always achieved.

Once the equivalent condensed tensors $\widehat{Y}_{ijrs}$ and $\widehat{S}_{pqrs}$ are determined, these can be exploited to define a class of effective second displacement gradient materials, through the following linear transformations

$$Y_{ijhkl} = Q_{hklrs}\widehat{Y}_{ijrs} + \Delta Y_{ijhkl},$$
$$S_{ijhklm} = Q_{ijhpq}Q_{klmrs}\widehat{S}_{pqrs} + \Delta S_{ijhklm}. \qquad (33)$$

Equivalence between the strain energy for the effective material and its 'condensed version' (31) implies

$$Y_{ijhkl} H_{ij} M_{hklpq}\widehat{\kappa}_{pq} = \widehat{Y}_{ijhk} H_{ij}\widehat{\kappa}_{hk},$$
$$S_{ijhklm} M_{ijhpq} M_{klmrs}\widehat{\kappa}_{pq}\widehat{\kappa}_{rs} = \widehat{S}_{ijhk}\widehat{\kappa}_{ij}\widehat{\kappa}_{hk}, \quad \forall H_{ij},\widehat{\kappa}_{ij}, \qquad (34)$$

so that the components of the tensor $\mathbf{Q}$ satisfy

$$M_{ijhrs}Q_{ijhpq} = \delta_{pr}\delta_{qs}, \qquad (35)$$

and the components of the tensors $\Delta\mathbf{Y}$ and $\Delta\mathbf{S}$ satisfy

$$\Delta Y_{ijhkl} M_{klmrs}\widehat{\kappa}_{rs} = 0,$$
$$\Delta S_{ijhklm} M_{klmrs}\widehat{\kappa}_{rs} = 0, \quad \forall\widehat{\kappa}_{rs}. \qquad (36)$$

It is evident that more than one triplet of tensors $\mathbf{Q}$, $\Delta\mathbf{Y}$ and $\Delta\mathbf{S}$ may satisfy the linear equations (35) and (36) so that, not one but, a class of pairs of effective non-condensed tensors gradient materials $\widehat{\mathbf{Y}}$ and $\widehat{\mathbf{S}}$ is found through the present homogenization scheme. More specifically, the components of the tensors $\Delta\mathbf{Y}$ and $\Delta\mathbf{S}$ satisfying Eq.(36) are given by the following relations

$$\Delta Y_{ijhkl} = L_{hklrst} A_{ijrst},$$
$$\Delta S_{ijhklm} = L_{ijhpqr} L_{klmstu} B_{pqrstu}, \qquad (37)$$

where $A_{ijrst}$ and $B_{pqrstu}$ remain undetermined from the present homogenization scheme, while

$$L_{ijhpqr} = \delta_{ip}\delta_{jq}\delta_{hr} - M_{ijhkl} R_{klst} M_{pqrst}. \qquad (38)$$

In Eq.(38), the components $R_{ijhk}$ are defined by the condition (solvable whenever the elasticity tensor $\mathbf{C}$ is positive definite)

$$R_{ijhk} M_{rsthk} M_{rstlm} = \delta_{il}\delta_{jm}, \qquad (39)$$

and satisfy the symmetry property $R_{ijhk} = R_{hkij}$, being its inverse a symmetric fourth-order tensor

$$R^{-1}_{ijhk} = M_{lmnij} M_{lmnhk}. \qquad (40)$$

The 'choice' of the tensor $\mathbf{Q}$ defines how the components of $\widehat{\mathbf{Y}}$ and $\widehat{\mathbf{S}}$ are mapped into the components of $\mathbf{Y}$ and $\mathbf{S}$. Among the infinite possibilities, the simple case where the obtained quantities correspond to a subset of the components of the non-condensed material is considered in the following. This is provided by the tensor $\mathbf{Q}$ with components

$$Q^*_{ijhpq} = \delta_{ip}\delta^{(q)}_{jh}, \qquad (41)$$

where $\delta_{jh}^{(q)}$ is the *three-index Kronecker delta* defined as

$$\delta_{jh}^{(q)} = \begin{cases} 1, & \text{if } q = j = h, \\ 0, & \text{otherwise.} \end{cases} \quad (42)$$

Another possible choice for the tensor $\mathbf{Q}$ (not considered in the following) corresponds to the minimization of the Frobenius norm of the tensors defining the equivalent non-condensed material and is represented by the tensor $\mathbf{Q}$ of components

$$Q^{**}_{ijhpq} = M_{ijhkl} R_{klpq}. \quad (43)$$

In any case, the properties of the condensed tensors $\widehat{\mathbf{Y}}$ and $\widehat{\mathbf{S}}$ imply in general that:

- when a centro-symmetric microstructure is considered, a null tensor $\Delta \mathbf{Y} = \mathbf{0}$ has to be enforced in order to guarantee a centro-symmetric response, $\mathbf{Y} = \mathbf{0}$;

- in the limit when the microstructure disappears and the material becomes homogeneous, the delta tensors have to be enforced to vanish as well, $\widehat{\mathbf{Y}} = \mathbf{0}$ and $\widehat{\mathbf{S}} = \mathbf{0}$, in order to turn back to the Cauchy behaviour, $\mathbf{Y} = \mathbf{S} = \mathbf{0}$;

- a positive semi-definite higher-order behaviour is achieved whenever $\Delta \mathbf{S} = \mathbf{0}$ is considered. Positive semi-definiteness may be turned into positive definiteness provided specific choices of $\Delta \mathbf{S} \neq \mathbf{0}$ are introduced, even small in modulus.

With respect to the last property, it is worth to remark that the tensor $\mathbf{Q}$ maps the four positive eigenvalues of the condensed tensor $\widehat{\mathbf{S}}$ into four positive eigenvalues of the (non-condensed) tensor $\mathbf{S}$. The remaining two eigenvalues of the tensor $\mathbf{S}$ are strictly related to the choice of the tensor $\Delta \mathbf{S}$.

*2.4. A technique for the evaluation of the perturbation functions*

The calculation of the perturbation functions $\mathbf{N}^{(1)}(\mathbf{x}/\epsilon)$ and $\widehat{\mathbf{N}}^{(2)}(\mathbf{x}/\epsilon)$ is the key step in the analytical evaluation of the effective tensors. To perform this calculation, the quadratic displacement field, Eq.(5), constrained by the equilibrium condition Eq.(9), is used as a boundary condition for the heterogeneous material (defined on the finite domain $\mathcal{L}$). The micro-displacement field $\mathbf{u}(\mathbf{x})$, solution of the elastic problem, can be expressed as

$$u_h(\mathbf{x}) = u_{hst}^\alpha(\mathbf{x}) \alpha_{st} + u_{hij}^\beta(\mathbf{x}) \widehat{\beta}_{ij}, \quad (44)$$

where $\mathbf{u}^\alpha(\mathbf{x})$ and $\mathbf{u}^\beta(\mathbf{x})$ are respectively the second- and third- order tensors, functions of the components of the tensors $\boldsymbol{\alpha}$ and $\boldsymbol{\beta}$, defining the imposed displacement boundary condition (5).

Restricting the attention to the unit central cell $\mathcal{A}$, a comparison between equation (44) and the asymptotic expansion equation (18) leads to the following identification of the perturbation functions

$$\epsilon N_{hst}^{(1)}\left(\frac{\mathbf{x}}{\epsilon}\right) = u_{hst}^\alpha(\mathbf{x}) - \delta_{hs} x_t, \quad (45a)$$

$$\epsilon^2 \widehat{N}_{hij}^{(2)}\left(\frac{\mathbf{x}}{\epsilon}\right) = u_{hij}^\beta(\mathbf{x}) + \left[\frac{1}{2}\delta_{hq} x_s x_t - u_{hqs}^\alpha(\mathbf{x}) x_t\right] M_{qstij}, \quad (45b)$$

two equations disclosing that $\mathbf{N}^{(1)}$ is independent of the transformation tensor $\mathbf{M}$.

Due to boundary layer effects, the obtained perturbation functions results to be only approximately periodic. However, it is expected that the periodicity of these functions is *asymptotically* recovered at increasing the repetition of the unit cell along both directions, which is to say, at large values of $n_1$ and $n_2$. In fact, it will be shown in Section 4 that periodicity is recovered with good approximation even for small repetitions of the unit cell.

### 3. Algorithms for the implementation of the finite element-based identification procedure

In order to perform the identification of the first- and second- order homogenized coefficients of the elastic constitutive tensors, the finite element method is exploited for solving the partial differential equations of elastic equilibrium for the heterogeneous material with any given period microstructure. Although the theoretical framework presented in the previous section is general and can be applied to two-dimensional or three-dimensional problems, the attention will be restricted for simplicity to plane stress or plane strain cases. Hence, considering a linear elastic body occupying the finite domain $\mathcal{L} \in \mathbb{R}^2$ chosen to be sufficiently large as compared to the elementary periodic cell, the standard isoparametric finite element discretization of the continuum has to be applied in order to move from the strong form of the governing equations to the corresponding weak form. Regarding the finite element topology, quadrilateral 8-nodes finite elements with quadratic shape functions are considered to provide a satisfactory approximation for the gradient of the perturbation functions which is required in the identification procedure. The same methodology applies to triangular quadratic finite elements. Numerical integration of the stiffness matrix and the force vector is performed using a 3×3 Gaussian quadrature formula, as usual.

Although the finite domain $\mathcal{L}$ is periodic in the $\mathbf{v}_1$ and $\mathbf{v}_2$ directions, the analysis is not now restricted to a single cell with periodic boundary conditions as is usually done in first-order computational homogenization schemes [4, 8, 22, 23, 27, 31, 43, 55]. More specifically, in classical computational homogenization techniques, the micro-displacement field is expressed as the superposition of an unknown micro-displacement fluctuation field to a macroscopic displacement field providing macro-strains assumed to be expressed by a polynomial. The unknown micro-displacement fluctuation field is obtained through the solution of homogeneous cell problems with vanishing body forces and prescribed periodic boundary conditions. In most of the second-order homogenization methods, the fluctuation field is assumed periodic and not all the second-order strain components can be controlled by such inhomogeneous periodic boundary conditions. To overcome such a problem, non-periodic and generalized periodic boundary conditions for second-order homogenization have been introduced [8, 27]. However, although the continuity of the macroscopic strain field is preserved in this case, the differentiability is in general not yet fully achieved, as further remarked in [8, 11].

Differently, the formulation proposed in this article (based on perturbation functions approaching periodicity and sufficient regularity at the boundary of the periodic cell by increasing the size of the cluster of cells) leads to differentiable macroscopic strain field. In this way, the obtained overall elastic tensors which become independent of the choice of the periodic cell. Therefore, the whole finite domain is modelled and a set of linear elastic problems characterized by different boundary conditions is solved. Such boundary conditions are given by Eq.(5), which combined with the kinematical constraint equation (15), and neglecting rigid-body motions ($\zeta_h = 0$ and $\alpha_{12} = \alpha_{21}$), become

$$u_h(\mathbf{x}) = \alpha_{hp} x_p + \frac{1}{2} M_{hpqij} \widehat{\beta}_{ij} x_p x_q, \qquad \mathbf{x} \in \partial \mathcal{L},$$

$$\text{with } \alpha_{12} = \alpha_{21}. \tag{46}$$

In particular, two classes of problems have to be solved. First, linear displacement boundary conditions are specified by considering the linear term of Eq.(46). This leads to three distinct boundary value problems $\mathfrak{P}_k^{(1)}$ ($k = 1, 2, 3$) with the following boundary conditions:

$$\mathfrak{P}_1^{(1)} : \begin{cases} u_1 = \alpha_{11} x_1, \\ u_2 = 0, \end{cases} \tag{47a}$$

$$\mathfrak{P}_2^{(1)} : \begin{cases} u_1 = 0, \\ u_2 = \alpha_{22} x_2, \end{cases} \tag{47b}$$

$$\mathfrak{P}_3^{(1)} : \begin{cases} u_1 = \alpha_{12} x_2, \\ u_2 = \alpha_{12} x_1. \end{cases} \tag{47c}$$

Due to the linearity in the mechanical response, any value of the parameters $\alpha_{11}$, $\alpha_{12}$, and $\alpha_{22}$ can be prescribed without affecting the computation of the perturbation function.

The solution of the above three boundary value problems leads to the displacement fields $u_{rhp}^\alpha(\mathbf{x}) \alpha_{hp}$ ($r = 1, 2$), representing the first term of Eq.(44). From the computed displacements of the finite elements belonging to the central cell $\mathcal{A}$ of the domain $\mathcal{L}$, the perturbation functions $N_{rhp}^{(1)}$ are determined as per Eq.(45a). This operation is performed through a record of the displacements evaluated at the Gauss points for each finite element of the cell $\mathcal{A}$ in an output file, ready for a post-processing operation.

After solving the set of boundary value problems $\mathfrak{P}_k^{(1)}$ ($k = 1, 2, 3$), another set of boundary value problems $\mathfrak{P}_k^{(2)}$ ($k = 1, ..., 4$) described by the following quadratic displacement boundary conditions is considered and solved:

$$\mathfrak{P}_1^{(2)} : \begin{cases} u_1 = \frac{1}{2} M_{11111} \widehat{\beta}_{11} x_1^2, \\ u_2 = \frac{1}{2} (M_{21211} + M_{22111}) \widehat{\beta}_{11} x_1 x_2, \end{cases} \tag{48a}$$

$$\mathfrak{P}_2^{(2)} : \begin{cases} u_1 = \frac{1}{2} (M_{11222} + M_{12122}) \widehat{\beta}_{22} x_1 x_2, \\ u_2 = \frac{1}{2} M_{22222} \widehat{\beta}_{22} x_2^2, \end{cases} \tag{48b}$$

$$\mathfrak{P}_3^{(2)} : \begin{cases} u_1 = \frac{1}{2} M_{12212} \widehat{\beta}_{12} x_2^2, \\ u_2 = \frac{1}{2} (M_{21212} + M_{22112}) \widehat{\beta}_{12} x_1 x_2, \end{cases} \tag{48c}$$

$$\mathfrak{P}_4^{(2)} : \begin{cases} u_1 = \frac{1}{2} (M_{11221} + M_{12121}) \widehat{\beta}_{21} x_1 x_2, \\ u_2 = \frac{1}{2} M_{21121} \widehat{\beta}_{21} x_1^2. \end{cases} \tag{48d}$$

The solution of the above four boundary value problems leads to the displacement fields $u_{rij}^\beta(\mathbf{x}) \widehat{\beta}_{ij}$ ($r = 1, 2$) which represent the second term of Eq.(44), so that as done previously for the three boundary value problems, from the computed displacements of the finite elements belonging to the central cell $\mathcal{A}$ of the domain $\mathcal{L}$, the perturbation functions $\widehat{N}_{rhp}^{(2)}$ are determined as per Eq.(19b), which does depend also on the displacement field solutions of the above $\mathfrak{P}_h^{(1)}$ boundary value problems with linear boundary conditions. Again, the computation is performed at each Gauss point level.

Once the perturbation functions $\mathbf{N}^{(1)}$ and $\widehat{\mathbf{N}}^{(2)}$ are determined, the six independent components $\mathcal{C}_{ijhk}(\boldsymbol{\xi})$ are computed for each Gauss point as functions of the microscopical constitutive elastic tensor and of the microscopic gradients of the perturbation function $\mathbf{N}^{(1)}$ at the same point, as per Eq.(26)$_1$. The computation of the microscopic gradients of the perturbation function $\mathbf{N}^{(1)}$ requires the determination of the partial derivatives of the microscopic displacement field components $u_{rhp}^\alpha$ ($r = 1, 2$) with respect to the $\xi_q$ ($q = 1, 2$) directions. For their accurate evaluation, it is preferable not to apply finite difference formulae to the computed displacements as a post-processing operation, but rather to compute them in the finite element routine via the partial derivatives of the shape element functions, which are not vanishing for quadratic finite elements. In formulae, the partial derivatives of the microscopic displacement field components read:

$$\frac{\partial u_{rhp}^\alpha}{\partial x_q} = \sum_{i=1}^{nen} \frac{\partial \mathcal{N}_i}{\partial x_q} u_{hrp}^\alpha, \quad (49)$$

where $\mathcal{N}_i(s_1, s_2)$ ($i = 1,...,nen$, with $nen$ the number of element nodes) is the standard finite element shape function dependent on the natural coordinates $s_1$ and $s_2$. For a 8-node quadrilateral finite element, $nen = 8$, $-1 \leq \{s_1, s_2\} \leq +1$ and the derivatives $\partial \mathcal{N}_i/\partial x_q$ have to be computed from the derivatives of the shape functions with respect to the natural coordinates $s_1$ and $s_2$, pre-multiplied by the inverse of the Jacobian matrix related to the transformation $(x_1, x_2) \rightarrow (s_1, s_2)$. Therefore, in conclusion, the microscopic gradients of the perturbation functions $\mathbf{N}^{(1)}$ are given as:

$$\frac{\partial N_{rhp}^{(1)}}{\partial \xi_q} = \left(\frac{\partial u_{rhp}^\alpha}{\partial x_q} - \delta_{rh}\delta_{pq}\right). \quad (50)$$

Next, the components $\widehat{\mathcal{Y}}_{ijhk}$ and $\widehat{\mathcal{S}}_{ijhk}$ are determined at the Gauss point level according to Eqs.(29). This requires the evaluation of the perturbation functions $\widehat{N}_{htij}^{(1)} = N_{hqs}^{(1)} M_{qstij}$ and the partial derivatives of $\widehat{\mathbf{N}}^{(2)}$ with respect to $\xi_t$ ($t = 1, 2$). The latter operation is performed in the post-processing routine as:

$$\frac{\partial \widehat{N}_{hij}^{(2)}}{\partial \xi_t} = \frac{1}{\epsilon}\left[\frac{\partial u_{hij}^\beta}{\partial x_t} + \left(\frac{1}{2}\delta_{hq}\delta_{st}x_r + \frac{1}{2}\delta_{hq}\delta_{tr}x_s\right.\right.$$
$$\left.\left. - \frac{\partial u_{hqs}^\alpha}{\partial x_t}x_r - u_{hqs}^\alpha \delta_{rt}\right)M_{qsrij}\right]. \quad (51)$$

Finally, the components of the overall constitutive tensors are determined according to Eq.(32). This is accomplished by a post-processing operation involving the numerical integration of the corresponding quantities over all the finite elements belonging to the central cell $\mathcal{A}$ of the cell-cluster $\mathcal{L}$:

$$C_{ijhk} = \frac{1}{\epsilon^2}\sum_{e=1}^{nfe}\sum_{g=1}^{ngp}\mathcal{C}_{ijhk}(\boldsymbol{\xi}_g)\det \mathbf{J}_g wgp_g, \quad (52a)$$

$$\widehat{Y}_{ijhk} = \frac{1}{\epsilon}\sum_{e=1}^{nfe}\sum_{g=1}^{ngp}\mathcal{Y}_{ijhk}(\boldsymbol{\xi}_g)\det \mathbf{J}_g wgp_g, \quad (52b)$$

$$\widehat{S}_{ijhk} = \sum_{e=1}^{nfe}\sum_{g=1}^{ngp}\mathcal{S}_{ijhk}(\boldsymbol{\xi}_g)\det \mathbf{J}_g wgp_g, \quad (52c)$$

where $g$ is the Gauss point number ranging from 1 to $ngp = 9$ for a 8-node quadrilateral finite element, $wgp_g$ is the integration weight, $\det \mathbf{J}_g$ is the determinant of the Jacobian matrix evaluated in the $g$-th Gauss point, and $nfe$ denotes the total number of finite elements belonging to the unit cell $\mathcal{Q}$.

The overall constitutive tensors of the second-order continuum are finally determined according to Eq.(33), together with the first of Eq.(32) and the choice for the tensor $\mathbf{Q} = \mathbf{Q}^*$ provided by Eq.(41).

## 4. Numerical examples

The proposed identification method is implemented and validated for two exemplary periodic micro-structures with different inclusion shape, namely, circular or quadrilateral, Fig. 2. In the latter case, different aspect ratios for the inclusion are examined, passing from square to rectangular inclusions. In all the presented cases, the boundary of the cell is a square with size $\epsilon$ (so that $r = 1$) and the two phases are modelled as linear elastic isotropic Cauchy material. In order to achieve reliable results from the presented algorithm, a preliminary analysis has to be conducted to determine the minimum size of the cluster of cells such that the perturbation functions for its central cell do not significantly vary by increasing the domain size. For its quantitative assessment, the maximum among the quadratic norms of the relative error for each perturbation function corresponding to two different domain sizes has been used as a control parameter. The performed computations have shown that clusters made up of $11 \times 11$ cells ($n_1 = n_2 = 11$) represent an excellent trade-off between accuracy and computation time for the solution of the linear elastic micro-scale heterogenous problem.

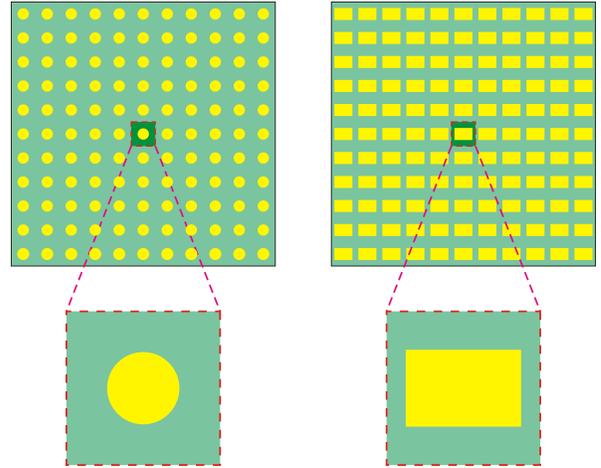

**Figure 2:** Two Cauchy composite materials obtained as the periodic repetition of square unit cells, containing a circular (left) or a quadrilateral (right) inclusion within the matrix.

### 4.1. Periodic cell containing a circular inclusion

A composite material with a periodic distribution of circular inclusions within a matrix is considered under plane strain conditions. The material is obtained as the repetition of a periodic square cell with side $\epsilon$, Fig. 2 (left). The isotropic matrix (characterized by Young's modulus $E_m = 60$ GPa and Poisson ratio $\nu_m = 0.3$) contains soft isotropic inclusions ($E_i = 3/17 E_m$, $\nu_i = 0.18$) for a volumetric content $f = 0.175$.

The application of the proposed approach to homogenization (Sec. 2) and algorithm (Sec. 3) leads to the definition of the perturbation function $\mathbf{N}^{(1)}$ which, due to the problem symmetry, has only three independent components $N_{111}$, $N_{211}$, and $N_{211}$, functions of the fast variable $\boldsymbol{\xi} = \mathbf{x}/\epsilon$, see Fig.3. It can be noted that the perturbation functions are found to be *approximately* $\mathcal{Q}$-periodic and regular on the border of the periodic cell, confirming the theoretical expectation. Once the perturbation function $\mathbf{N}^{(1)}$ is determined, the local overall elastic constitutive tensor $\mathbf{C}$ can be computed, which results to possess *cubic* symmetry and the following non vanishing components

$$C_{1111} = C_{2222} = 5.75 \times 10^4 \,\text{MPa},$$
$$C_{1122} = C_{2211} = 2.15 \times 10^4 \,\text{MPa}, \quad (53)$$
$$C_{1212} = 1.68 \times 10^4 \,\text{MPa}.$$

The perturbation function $\widehat{\mathbf{N}}^{(2)}$ is determined using $\mathbf{C}$ and $\mathbf{N}^{(1)}$ according to the procedure detailed in Sec. 3 and, similarly to the perturbation function $\mathbf{N}^{(1)}$, it can be noted from a selection of the components of $\widehat{\mathbf{N}}^{(2)}$ shown in Fig.4 to be *approximately* periodic and regular on the boundary of the periodic cell.

Finally, the *condensed* nonlocal overall elastic constitutive tensors $\widehat{\mathbf{Y}}$ and $\widehat{\mathbf{S}}$ are determined. Due to the centro-symmetry of the considered cell geometry, the coupling tensor vanishes, $\widehat{\mathbf{Y}} = 0$, while the $\widehat{\mathbf{S}}$ tensor has the following non vanishing components:

$$\widehat{S}_{1111}/\epsilon^2 = \widehat{S}_{2222}/\epsilon^2 = 900.0 \,\text{MPa},$$
$$\widehat{S}_{1212}/\epsilon^2 = \widehat{S}_{2121}/\epsilon^2 = 54.2 \,\text{MPa}, \quad (54)$$
$$\widehat{S}_{1112}/\epsilon^2 = -\widehat{S}_{2221}/\epsilon^2 = 25.4 \,\text{MPa},$$

and its eigenvalues are given by $\widehat{\sigma}_1/\epsilon^2 = \widehat{\sigma}_2/\epsilon^2 = 904.8$ MPa and $\widehat{\sigma}_3/\epsilon^2 = \widehat{\sigma}_4/\epsilon^2 = 53.4$ MPa, so that the nonlocal overall elastic constitutive tensor $\widehat{\mathbf{S}}$ is positive definite. Moreover, the constitutive higher-order response is also characterized by cubic symmetry.

Considering the obtained *condensed* tensors $\widehat{\mathbf{Y}}$ and $\widehat{\mathbf{S}}$, the nonlocal overall elastic constitutive tensors $\mathbf{Y}$ and $\mathbf{S}$ can be evaluated from Eq. (33), once the structure for the fifth-order tensor $\mathbf{Q}$ is specified together with that for $\Delta \mathbf{Y}$ and $\Delta \mathbf{S}$. Restricting the attention to the case $\Delta \mathbf{Y} = \Delta \mathbf{S} = 0$, and $\mathbf{Q} = \mathbf{Q}^*$, Eq.(41), the coupling tensor vanishes, $\mathbf{Y} = 0$, while the $\mathbf{S}$ tensor has the following non-null components

$$S_{111111}/\epsilon^2 = S_{222222}/\epsilon^2 = 900.0 \,\text{MPa},$$
$$S_{122122}/\epsilon^2 = S_{211211}/\epsilon^2 = 54.2 \,\text{MPa}, \quad (55)$$
$$S_{111122}/\epsilon^2 = -S_{222211}/\epsilon^2 = 25.4 \,\text{MPa}.$$

To assess the validity of the obtained effective non local models, the mechanical response of the heterogenous Cauchy material, obtained through finite element simulations, is compared with that of the homogenized second-order continuum for two benchmark boundary value problems. The two boundary value problems correspond to the cases of periodic body-forces and of simple shear, for which the response of the homogenized second-order material is one-dimensional and described by an analytical closed form solution.

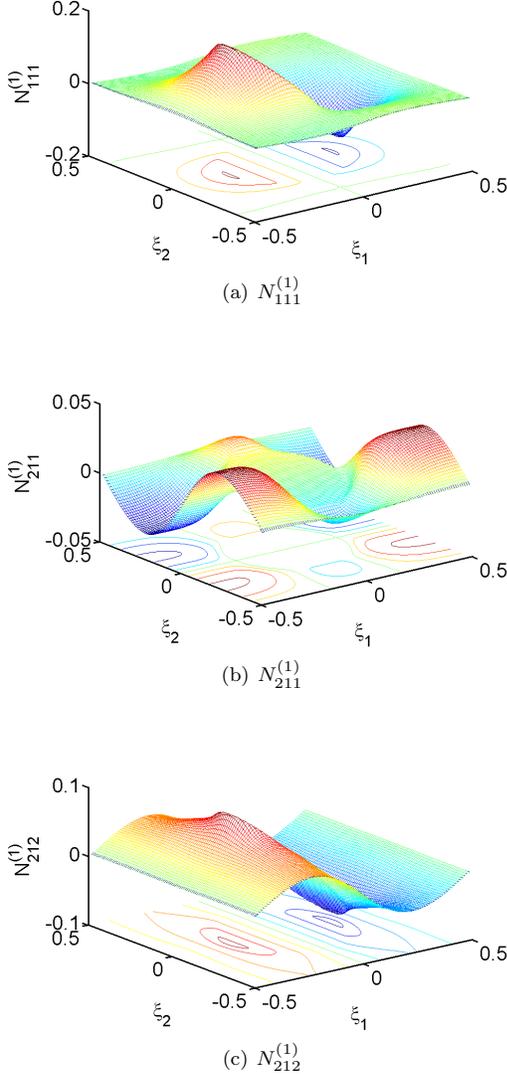

**Figure 3:** Components of the perturbation function $\mathbf{N}^{(1)}$ vs. $\xi_1 = x_1/\epsilon$ and $\xi_2 = x_2/\epsilon$ for the circular inclusion problem.

*First benchmark test.* An infinite periodic composite is subject to a $\mathcal{L}-$periodic body force defined as a sinusoidal

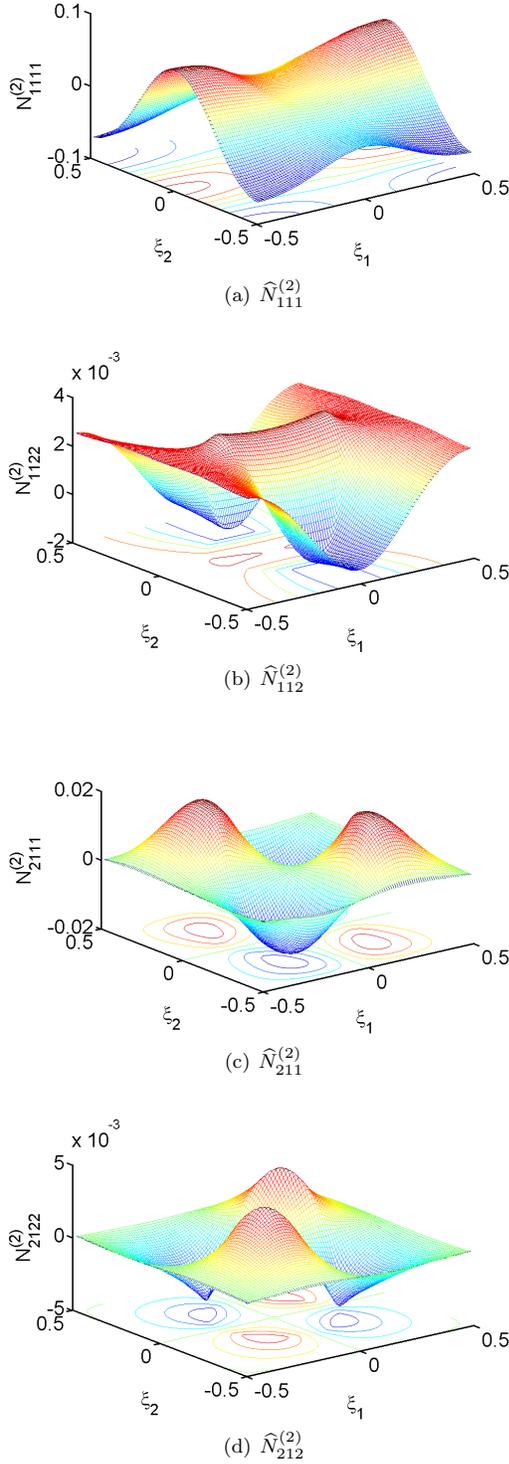

**Figure 4:** A selection of the components of the perturbation function $\widehat{\mathbf{N}}^{(2)}$ vs. $\xi_1 = x_1/\epsilon$ and $\xi_2 = x_2/\epsilon$ for the circular inclusion problem.

force acting only parallel to the $x_1$-axis

$$f_1(x_2) = \Xi \sin\left(\frac{2\pi x_2}{L}\right), \qquad f_2 = 0, \qquad (56)$$

where $\Xi$ represents the body-force amplitude while $L$ is the distance measuring the half of the body-force periodicity. Under this loading condition, the solution for the macro-displacement is given by

$$U_1(x_2) = \left(\frac{L}{2\pi}\right)^2 \frac{\Xi \sin\left(\frac{2\pi x_2}{L}\right)}{C_{1212}\left[1 + \left(2\pi\frac{\lambda_1^{\text{sh}}}{L}\right)^2\right]}, \qquad (57)$$

where the characteristic lengthscale $\lambda_1^{\text{sh}}$ in shear has been introduced in the form

$$\lambda_1^{\text{sh}} = \sqrt{\frac{S_{122122}}{C_{1212}}}. \qquad (58)$$

Assuming that the periodicity of the body force equals $n_2$ times the characteristic size $\epsilon$ of the periodic cell ($L = n_2\epsilon$), the response of the heterogenous material was simulated by imposing standard periodic boundary conditions to a layer made up of a stack of $n_2$ square cells aligned parallel to the $x_2$-axis. The macroscopic displacement and displacement gradient related to the solution of the microscopic problem are finally computed according to the up-scaling relations evaluated through an average (over the microscopic variable $\boldsymbol{\xi}$) of the corresponding microscopic fields defined over the periodic cell (for details on up-scaling relations see [8, 11, 48]. With reference to the considered cell-cluster ($n_2 = 11$), the characteristic shear length (58) results to be

$$\lambda_1^{\text{sh}} = 0.057\epsilon, \qquad (59)$$

so that the comparison between the two responses is reported in Fig.5 (upper part) in terms of the gradient of the macro-displacement component $H_{12}$ (made dimensionless through multiplication by $2\pi C_{1212}/(\Xi L)$). An excellent agreement is shown between the analytical solution for the second-order continuum based on the identified homogenized elastic parameters (continuous red line) and the corresponding response of the heterogenous problem obtained through finite element simulation (diamond spots).

*Second benchmark test.* An infinite layer made up of a stack of rectangular cells, with height $L = n_2\epsilon$, aligned parallel to the $x_1$-axis is subject to a simple shear displacement, corresponding to the following boundary conditions

$$\begin{aligned} U_1(x_2 = L) &= \overline{U}, \\ U_1(x_2 = 0) &= U_2(x_2 = 0) = U_2(x_2 = L) = 0, \end{aligned} \qquad (60)$$

and null body forces. The numerical results obtained for the considered heterogeneous material are compared with the analytical prediction for the effective higher-order material when the boundary conditions are complemented by

$$U_{1,1}(x_2 = 0) = U_{1,1}(x_2 = L) = 0. \tag{61}$$

The analytical solution for the displacement field in the homogeneous higher-order material is given by

$$U_1(x_2) = \overline{U}\left\{\left[1 - \cosh\left(\frac{L}{\lambda_1^{sh}}\right)\right]\left[1 - \cosh\left(\frac{x_2}{\lambda_1^{sh}}\right)\right]\right.$$
$$\left. + \sinh\left(\frac{L}{\lambda_1^{sh}}\right)\left[\frac{x_2}{\lambda_1^{sh}} - \sinh\left(\frac{x_2}{\lambda_1^{sh}}\right)\right]\right\}$$
$$\times \left\{2\left[1 - \cosh\left(\frac{L}{\lambda_1^{sh}}\right)\right] + \frac{L}{\lambda_1^{sh}}\sinh\left(\frac{L}{\lambda_1^{sh}}\right)\right\}^{-1}. \tag{62}$$

The resultant shear stress defined for higher-order materials [32, 44]

$$T_{12} = C_{1212}U_{1,2} - S_{122122}U_{1,222}, \tag{63}$$

is provided by the analytical solution as

$$T_{12} = KC_{1212}\frac{\overline{U}}{L}, \tag{64}$$

where $K$ represents an amplification stress factor due to the presence of the non locality and is given by

$$K = \frac{1}{1 - \frac{2\lambda_1^{sh}}{L}\tanh\left(\frac{L}{2\lambda_1^{sh}}\right)} \geq 1. \tag{65}$$

The comparison between the responses of the effective higher-order model and of the heterogenous material is shown in Fig.5 (lower part) in terms of the macro-rotation $\Omega_{21} = (U_{2,1} - U_{1,2})/2$ multiplied by the factor $2L/\overline{U}$. The continuous red line (analytical solution for the second-order continuum based on the identified homogenized elastic parameters) is in excellent agreement with the diamond spots (which represent the finite element predictions for the heterogenous problem), thus confirming once more the accuracy of the proposed model.

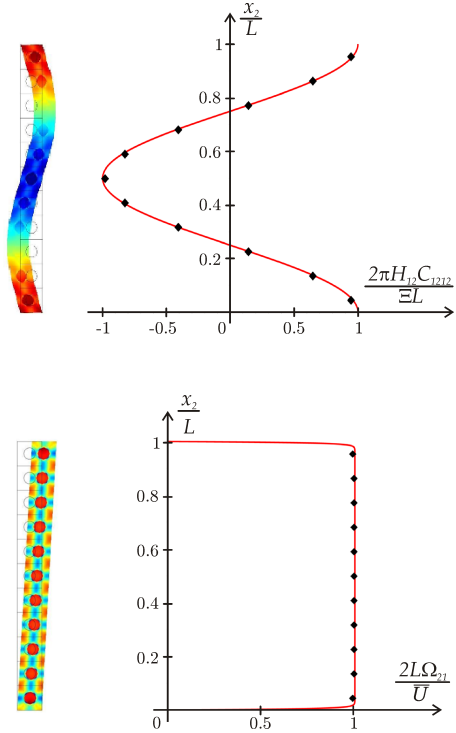

**Figure 5:** Comparison between the analytical solution for the second-order homogenized continuum based on the identified elastic parameters (red lines) and the finite element prediction (diamond spots) for the heterogenous problem: first benchmark (upper part) and second benchmark (lower part) tests.

### 4.2. Periodic cell containing a quadrilateral inclusion

A composite material with square unit cell containing a quadrilateral inclusion is considered under plane stress conditions. The matrix is assumed to be made up of an isotropic linear elastic material ($E_m = 100$ GPa, $\nu_m = 0.3$) containing an isotropic linear elastic inclusion, much softer than the matrix ($E_i = E_m/1000$ and $\nu_i = 0.3$). Different values of the volumetric fraction $f$ are examined, exploring the range from $f = 0$ to $f = 0.48$, within which the aspect ratio of the inclusion varies, Fig. 6. For $f$ ranging between 0 and 0.25, the inclusion is assumed to have a squared shape and its side is varied from 0 to $\epsilon/2$. For $f$ ranging between 0.25 and 0.48, the inclusion shape is modified into a rectangle by keeping one side constant and equal to $\epsilon/2$, while the other side parallel is increased up to $0.96\epsilon$.

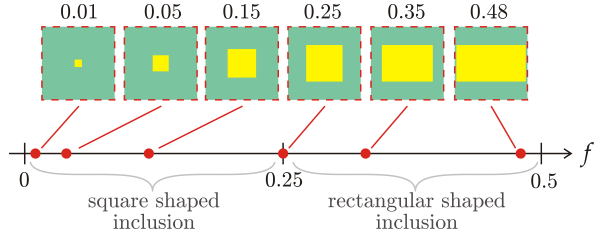

**Figure 6:** Shape of the quadrilateral inclusion at varying the volume content $f$ assumed in the numerical examples.

The application of the proposed theory and the algorithm detailed in Sec. 3 is repeated here as for the previous example to determine the perturbation function $\mathbf{N}^{(1)}$. A selection of the obtained perturbation functions, namely $N_{111}^{(1)}$, $N_{112}^{(1)}$, $N_{122}^{(1)}$ and $N_{222}^{(1)}$, are shown in Fig.7 for the case $f = 0.4$ where the largest side of the rectangular soft inclusion is parallel to the $x_1$ axis. It has to be remarked that the perturbation functions are approximately $Q-$periodic and regular on the border of the periodic cell, confirming the theoretical expectation.

Once the perturbation function $\mathbf{N}^{(1)}$ is determined, the local overall elastic constitutive tensor $\mathbf{C}$ can be computed. For the case $f = 0.4$ where the largest side of the rectangular soft inclusion is parallel to the $x_1$ axis, the non vanishing components of $\mathbf{C}$ are

$$\begin{aligned}
C_{1111} &= 5.17 \times 10^4 \, \text{MPa}, \\
C_{2222} &= 2.49 \times 10^4 \, \text{MPa}, \\
C_{1122} &= C_{2211} = 4.55 \times 10^3 \, \text{MPa}, \\
C_{1212} &= 2.41 \times 10^3 \, \text{MPa},
\end{aligned} \quad (66)$$

so that *orthotropic* symmetry emerges in the local effective response.

At this stage, the perturbation function $\mathbf{N}^{(2)}$ is determined using $\mathbf{C}$ and $\mathbf{N}^{(1)}$ and according to the procedure detailed in the previous section. Again, the perturbation function $\mathbf{N}^{(2)}$ is approximately periodic and regular on the edges of the periodic cell, as it can be seen from a selection of the components shown in Fig.8 for the case $f = 0.4$ and the considered orientation of the rectangular inclusion.

The *condensed* nonlocal overall elastic constitutive tensors $\widehat{\mathbf{Y}}$ and $\widehat{\mathbf{S}}$ are determined. For the considered geometry, the condensed tensor $\widehat{\mathbf{S}}$ has the following non-null compo-

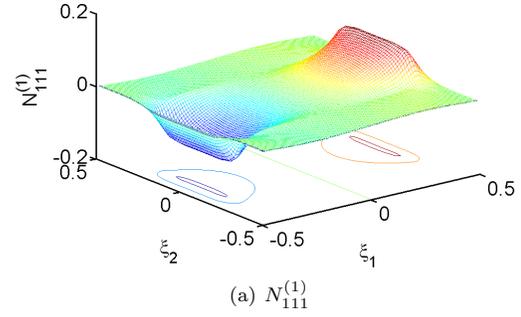
(a) $N_{111}^{(1)}$

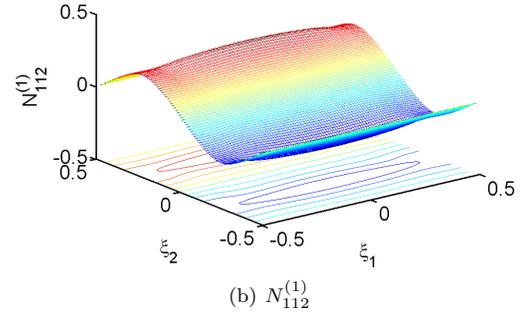
(b) $N_{112}^{(1)}$

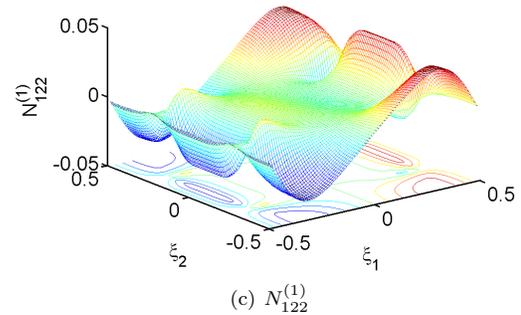
(c) $N_{122}^{(1)}$

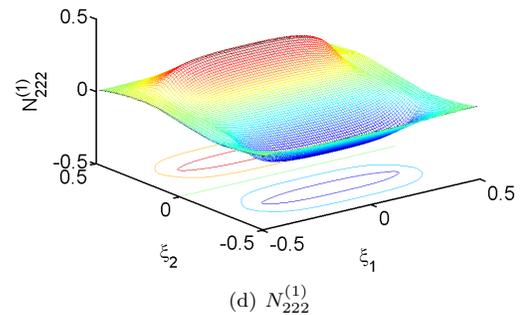
(d) $N_{222}^{(1)}$

**Figure 7:** Components of the perturbation function $\mathbf{N}^{(1)}$ for the rectangular inclusion problem when $f = 0.4$.

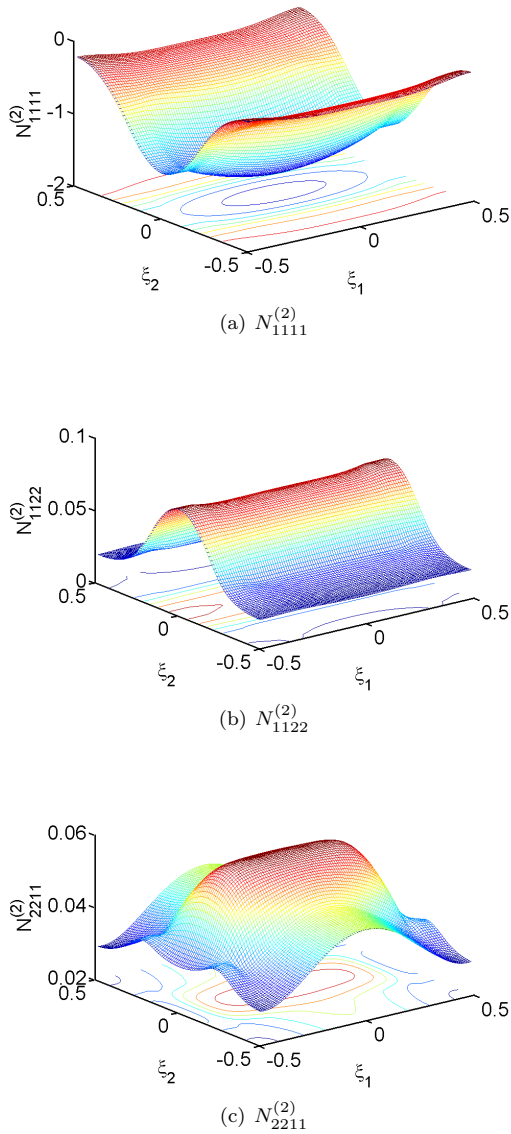

Figure 8: A selection of the components of the perturbation function $\mathbf{N}^{(2)}$ for the rectangular inclusion problem when $f = 0.4$.

nents

$$\widehat{S}_{1111}/\epsilon^2 = 1.4 \times 10^4 \, \text{MPa},$$
$$\widehat{S}_{2222}/\epsilon^2 = 1.5 \times 10^4 \, \text{MPa},$$
$$\widehat{S}_{1212}/\epsilon^2 = 54.6 \, \text{MPa},$$
$$\widehat{S}_{2121}/\epsilon^2 = 4.9 \times 10^2 \, \text{MPa},$$
$$\widehat{S}_{1112}/\epsilon^2 = -7.0 \times 10^2 \, \text{MPa},$$
$$\widehat{S}_{2221}/\epsilon^2 = -1.4 \times 10^3 \, \text{MPa},$$

while the coupling tensor vanishes, $\widehat{\mathbf{Y}} = 0$, because of centro-symmetric properties of the unit cell. Assuming $\Delta \mathbf{Y} = \Delta \mathbf{S} = 0$ and $\mathbf{Q} = \mathbf{Q}^*$, Eq.(41), the coupling tensor is null, $\mathbf{Y} = 0$, while tensor $\mathbf{S}$ has the following non-null components

$$S_{111111}/\epsilon^2 = 1.4 \times 10^4 \, \text{MPa},$$
$$S_{222222}/\epsilon^2 = 1.5 \times 10^4 \, \text{MPa},$$
$$S_{122122}/\epsilon^2 = 54.6 \, \text{MPa},$$
$$S_{211211}/\epsilon^2 = 4.9 \times 10^2 \, \text{MPa},$$
$$S_{111122}/\epsilon^2 = -7.0 \times 10^2 \, \text{MPa},$$
$$S_{222211}/\epsilon^2 = -1.4 \times 10^3 \, \text{MPa}.$$

*Benchmark tests.* The effectiveness of the obtained equivalent properties is assessed by comparing the mechanical response of the higher-order material with that of (i.) the heterogeneous material (predicted through finite element simulations). The comparison is also extended to the predictions obtained by applying (ii.) the second-order computational two-scale homogenization model (based on the well-known 'FE2 method' [27, 24, 30, 35, 36, 52]) for the estimation of the overall constitutive tensors and (iii.) the second-order asymptotic homogenization method considering the simplifying assumption of a quadratic macroscopic displacement field (see [8, 11]). Simple shear loading condition (as described in the previous subsection) is considered as a benchmark test. The comparison in terms of the parameter $K$, Eq.(65), for the different homogenization schemes is graphically shown in Fig.9 for two orthogonal orientations of the quadrilateral inclusion defining the composite material. In particular, the two orientations correspond to the shortest (Fig.9, upper part) and longest (Fig.9, lower part) side of the quadrilateral inclusion parallel to the simple shear direction, $x_1$–axis. The prediction (red solid line) of the presented model is in good agreement with the heterogenous finite element solution (blue solid line) and the asymptotic homogenization prediction (orange spots). The prediction of the method based on the FE2 approach (green solid line), on the other hand, is less accurate than the other methods.

## 5. Conclusions

An identification scheme has been proposed for higher-order elastic material equivalent to a heterogeneous (periodic) Cauchy elastic composite. The scheme is based on results from asymptotic homogenization, but is simple in the implementation. Results have been shown to provide a reasonably good accuracy in the description of the homogenized response of the composite material.

## Acknowledgments

AB and MP gratefully acknowledge financial support from the European Research Council to the ERC Starting Grant 'Multi-field and multi-scale Computational Approach to Design and Durability of PhotoVoltaic Modules' (ERC Grant Agreement n. 306622 - CA2PVM). FDC and DB gratefully acknowledge financial support from the European Research Council to the ERC Advanced Grant 'Instabilities and nonlocal multiscale modelling of materials' (ERC Grant Agreement n. 340561 - INSTABILITIES).

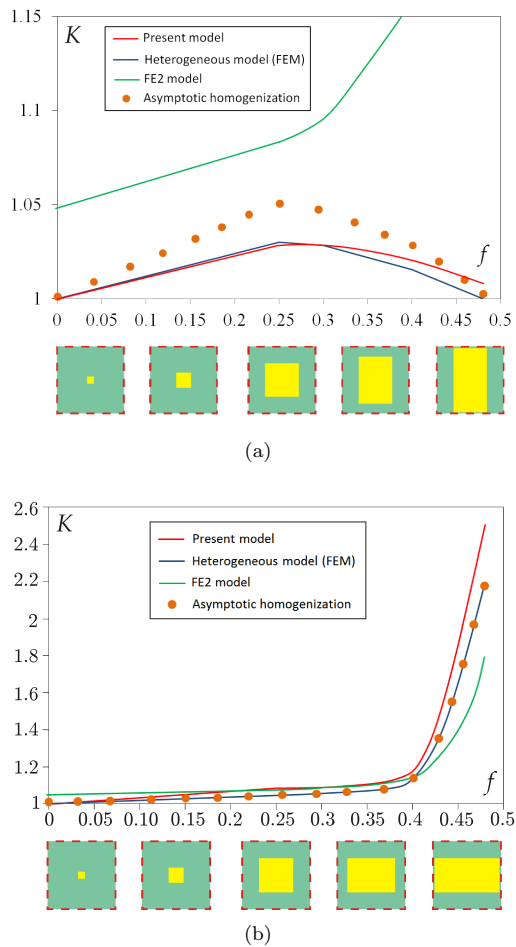

**Figure 9:** Comparison between different homogenization schemes for simple shear displacement conditions applied to a stack of quadrilateral inclusions. The case is considered where the quadrilateral inclusion has the shortest (a) or the longest (b) side parallel to simple shear direction.